%% file: Gd_ruler_paper.tex
\documentclass[reprint,superscriptaddress,amsmath,amssymb,aps,pra,]{revtex4-1}

\usepackage{graphicx}
\usepackage{dcolumn}
\usepackage{bm}
\usepackage[version=3]{mhchem}

\begin{document}

\title{Gd$^{3+}$ - Gd$^{3+}$ distances exceeding 3 nm determined by very high frequency continuous wave electron paramagnetic resonance}
\thanks{Electronic Supplementary Information (ESI) available: Additional sample details are provided, including the calculated most probable Gd-Gd distances of the Gd-rulers at the relevant temperatures. CW EPR spectra of Gd-4-iodo-PyMTA and the Gd-rulers recorded at 240 GHz and 30 K, 215 K, and 288 K, W-band echo-detected EPR of Gd-4-iodo-PyMTA and Gd-rulers at 10 K, and CW EPR spectra of Gd-DOTAM and Gd-NO3Pic recorded at 240 GHz and 30 K are provided. The syntheses of Gd-4-iodo-PyMTA and the Gd-rulers $\mathbf{2_{1}}$ and $\mathbf{2_{2}}$ are described and the NMR spectra of all compounds whose syntheses are described are shown.}%

\author{Jessica A. Clayton}
\affiliation{Department of Physics, University of California, Santa Barbara, Santa Barbara, CA, USA.}
\affiliation{Institute for Terahertz Science and Technology, University of California, Santa Barbara, Santa Barbara, CA, USA.}

\author{Mian Qi}
\affiliation{Faculty of Chemistry and Center for Molecular Materials (CM2), Bielefeld University, Bielefeld, Germany.}

\author{Adelheid Godt}
\affiliation{Faculty of Chemistry and Center for Molecular Materials (CM2), Bielefeld University, Bielefeld, Germany.}

\author{Daniella Goldfarb}
\affiliation{Department of Chemical Physics, Weizmann Institute of Science, Rehovot, Israel.}

\author{Songi Han}
\affiliation{Institute for Terahertz Science and Technology, University of California, Santa Barbara, Santa Barbara, CA, USA.}
\affiliation{Department of Chemistry and Biochemistry, University of California, Santa Barbara, Santa Barbara, CA, USA.}
\affiliation{Department of Chemical Engineering, University of California, Santa Barbara, Santa Barbara, CA, USA.}

\author{Mark S. Sherwin}
\affiliation{Department of Physics, University of California, Santa Barbara, Santa Barbara, CA, USA.}
\affiliation{Institute for Terahertz Science and Technology, University of California, Santa Barbara, Santa Barbara, CA, USA.}%

\date{\today}

\begin{abstract}
Electron paramagnetic resonance spectroscopy in combination with site-directed spin-labeling is a very powerful tool for elucidating the structure and organization of biomolecules. \ce{Gd^{3+}} complexes have recently emerged as a new class of spin labels for distance determination by pulsed EPR spectroscopy at Q- and W-band. We present CW EPR measurements at 240 GHz (8.6 Tesla) on a series of Gd-rulers of the type Gd-PyMTA---spacer---Gd-PyMTA, with Gd-Gd distances ranging from 1.2 nm to 4.3 nm. CW EPR measurements of these Gd-rulers show that significant dipolar broadening of the central $|-1/2\rangle\rightarrow|1/2\rangle$ transition occurs at 30 K for Gd-Gd distances up to $\sim$ 3.4 nm with Gd-PyMTA as the spin label. This represents a significant extension for distances accessible by CW EPR, as nitroxide-based spin labels at X-band frequencies can typically only access distances up to $\sim$ 2 nm. We show that this broadening persists at biologically relevant temperatures above 200 K, and that this method is further extendable up to room temperature by immobilizing the sample in glassy trehalose. We show that the peak-to-peak broadening of the central transition follows the expected $~1/r^3$ dependence for the electron-electron dipolar interaction, from cryogenic temperatures up to room temperature. A simple procedure for simulating the dependence of the lineshape on interspin distance is presented, in which the broadening of the central transition is modeled as an $S=1/2$ spin whose CW EPR lineshape is broadened through electron-electron dipolar interactions with a neighboring $S=7/2$ spin. 
\end{abstract}

\maketitle

\section{\label{sec:level1}Introduction}

Electron paramagnetic resonance (EPR) techniques in combination with site-directed spin labeling (SDSL) have become essential tools in structure studies of biomolecules. EPR methods - including pulsed dipolar EPR spectroscopy (PDS) and continuous wave (CW) EPR spectroscopy - complement conventional high-resolution structure measurements, such as X-ray crystallography and NMR spectroscopy. EPR spectrosopy is particularly advantageous in the study of complex protein-lipid assemblies where practical difficulties with low-yield expression and crystallization often preclude other biophysical characterization techniques.\cite{HemmingaM.A.L.J.Berliner2007} The determination of nanometer-scale distances between spin labeled sites in a biomolecule via EPR spectroscopy can provide information on the structure and organization of biomolecules, and can also be used for the tracking of conformational changes.\cite{Stone1965,Hubbell1994,Hustedt1999,Hubbell2000,Borbat2001} Spin labeling for EPR has been dominated by the use of nitroxide radicals, placed at carefully selected sites to generate a system with two proximal electron spins - either by doubly spin labeling or by multimer formation of singly spin labeled biomolecules.\cite{Shelke2011,Steinhoff2004} The distance between two such spin labels may be determined via PDS at X-band frequencies ($\sim$ 9.5 GHz), which can detect dipolar interactions between nitroxide radicals up to 8.0 nm apart, but has limited utility below 2.0 nm.\cite{Schiemann2007,Jeschke2007,Tsvetkov2008} PDS is complemented by CW EPR lineshape analysis, which allows for the determination of distances in the range of approximately 0.7 - 2.0 nm.\cite{Hubbell1998} CW EPR finds particular utility in the study of membrane proteins, offering a site-specific probe of structure and dynamics in native or native-mimicking environments under ambient solution conditions. However, distances in the borderline region of applicability of CW EPR and PDS - particularly in the 1.6 - 1.9 nm range - still remain difficult to access using spin labels based on nitroxide radicals.\cite{Banham2008,Borbat2014} Given the importance of discerning the structure and structural changes in biomolecules within this distance regime, spin labels that can fill in this gap are needed. Furthermore, CW EPR methods have greater signal sensitivity than PDS and tend to be applicable at higher - including ambient - temperatures, unlike PDS that typically must be applied at cryogenic temperatures.\cite{Jeschke2007} Thus, extending the distance range accessible by CW EPR methods beyond 2.0 nm is a highly desirable goal.

Increasing availability of high-frequency microwave sources and components has allowed routine EPR measurements to move from X-band ($\sim$ 9.5 GHz) and Q-band ($\sim$ 35 GHz), up to W-band ($\sim$ 95 GHz) and higher frequencies, including G-band (110 - 300 GHz) where sensitivity is increased and faster motional dynamics may be accessed.\cite{Schweiger2001,Polyhach1076,Ghimire2009,Goldfarb2008,Cruickshank2009,Smith1998,Tkach2014} While nitroxide-based spin labels are very suitable at the lower microwave frequencies, at frequencies above Q-band the performance of nitroxide labels for CW EPR distance measurement wanes due to strong inhomogeneous line broadening. During the last 10 years, Gd-based spin labels have become an important tool for structure studies with EPR at Q-band frequencies and above.\cite{Razzaghi2014,Gordon-Grossman2011,Garbuio2013,Matalon2013,Manukovsky,Song2011,Potapov2010,Yagi2011,Qi2014,Martorana2014,Theillet2016,Mascali2016} High-frequency EPR applications relying on \ce{Gd^{3+}} complexes have been demonstrated in peptides,\cite{Gordon-Grossman2011,Garbuio2013,Matalon2013,Manukovsky} nucleic acids,\cite{Song2011} proteins\cite{Potapov2010,Yagi2011} and in in-cell environments.\cite{Qi2014,Martorana2014,Theillet2016,Mascali2016} Gd-based spin labels possess several distinct advantages over nitroxide-based spin labels for EPR at high fields. These benefits include (i) a high-spin $S=7/2$ ground state, which confers longer range dipolar interactions than an $S=1/2$ system and thus increases the accessible distance range in CW EPR measurements, (ii) an isotropic \textit{g}-value, (iii) a sharp central $|-1/2\rangle\rightarrow|1/2\rangle$ transition that narrows with increasing magnetic field, and (iv) no orientation selection, the latter three of which concentrate the distance information into a narrow single resonance, resulting in increased sensitivity. Furthermore, Gd-based spin labels have been shown to be much more stable than nitroxide radicals in reducing environments, allowing for in-cell applications.\cite{Martorana2014,Qi2014,Mascali2016,Theillet2016} 

For $D\ll g\mu_{B}B_{0}$, a condition that is generally satisfied by existing \ce{Gd^{3+}} complexes at Ka-band and higher frequencies, the $|-1/2\rangle\rightarrow|1/2\rangle$ transition appears as an intense central peak in the EPR spectrum on top of a broad background due to all other transitions.\cite{Raitsimring2005} The width of this central transition scales as $D^{2}/g\mu_{B}B_{0}$, leading to a very narrow linewidth and thus conferring greater sensitivity to dipolar broadening of the CW EPR lineshape at high magnetic fields as compared to lower magnetic fields where the central transition is prohibitively broad.\cite{Goldfarb2014,Potapov2010} As a result of this narrow linewidth, it is possible to resolve the contribution of the dipolar interaction with a proximal \ce{Gd^{3+}} species to the lineshape of the CW EPR spectrum, manifested as an increase of the peak-to-peak linewidth of the $|-1/2\rangle\rightarrow|1/2\rangle$ transition. This was recently shown by Edwards et al. in frozen glassy solutions of \ce{GdCl3} in \ce{D2O}/\ce{glycerol-d8} at 240 GHz and 10 K.\cite{Edwards2013} Average interspin distances were manipulated by varying the concentration of \ce{GdCl3} from 50 mM to 0.1 mM, yielding corresponding mean interspin distances ranging from 1.8 nm to 6.6 nm. Substantial dipolar broadening was observed up to a mean interspin distance of 3.8 nm which is about twice the longest interspin distance resolvable by CW EPR lineshape analysis of nitroxide radicals at X-band. This is an important prospect, given that PDS using Gd-based spin labels is particularly prone to artifacts below 3 nm.\cite{Dalaloyan2015,Cohen2016} 

Extending the distance measurement sensitivity to above 3 nm would be an important milestone also because this length scale spans many relevant intra-protein distances, as well as many inter-protomer distances in protein oligomers. For a sense of scale, consider a pair of nitroxide-based spin labels at X-band, which can easily give access to the information on an interspin distance of 1.5 nm, and ask what fraction of a typical protein can be probed. Most proteins fold into globular domains, consisting of tightly packed atoms with an approximate density of 1.37 g/cm$^{3}$. A spherical region of mass M within such a globular protein has a radius $R(\textrm{nm})=0.6\ast M^{1/3}$, where the mass is given in Daltons.\cite{Erickson2009} A sphere of radius $R=1.5$ nm encloses a protein mass of 18 kDa, encompassing approximately 160 amino acids that could potentially be spin labeled. Now consider doubling this radius to 3 nm, a distance easily accessible by Gd-based spin labels at 240 GHz. This yields a protein mass of 94 kDa containing 850 amino acids - nearly eight times as many residues in the protein that can potentially be spin labeled for the measurement of intra-protein distances. This increased flexibility in choosing sites for mutagenesis and spin labeling is particularly useful in cases where e.g. the structure of a protein or protein complex is little known and therefore the distance estimated from a model of the protein may be different from the real distance, or in cases where mutation of certain amino acids is not possible due to the nature of the mutation (e.g. charged amino acids) or is limited by geometric restriction of the site.

In this paper, we present a fundamental study of the CW EPR lineshape analysis-based approach to determining Gd-Gd distances at very high frequencies (240 GHz). We rely on a series of compounds of the type Gd-PyMTA---spacer---Gd-PyMTA (Fig. 1) which we call Gd-rulers, to test the validity of this approach. The rod-like spacer keeps the two Gd-based spin labels, Gd-PyMTA, and hence the two \ce{Gd^{3+}} ions at a well-defined distance. Experiments with these Gd-rulers show that dipolar broadening of the central transition of \ce{Gd^{3+}} is detectable in such a biradical system at cryogenic temperatures, and that the maximal distance sensitivity is consistent with that estimated from the study with random solutions of \ce{GdCl3}. A simple procedure for simulating the dependence of \ce{Gd^{3+}} lineshape on interspin distance is described, in which the broadening of the central transition is modeled as an $S=1/2$ spin whose CW EPR lineshape is broadened through electron-electron dipolar interactions with a neighboring $S=7/2$ spin. The well-known spacer stiffness, and therefore the calculable most probable Gd-Gd distances $r$ of the Gd-rulers \cite{Jeschke2010} allow for a careful check on the relationship between the measured line broadening and the distance $r$, revealing that the peak-to-peak broadening follows the expected $1/r^{3}$ dependence of the dipolar interaction. Furthermore, we show that this correlation is maintained even at biologically relevant temperatures - from around the protein dynamical transition temperature ($\sim$ 190 - 220 K)\cite{Doster2010} up to room temperature for samples immobilized in a glassy matrix.

\section{\label{sec:level1}Experimental}

\subsection{\label{sec:level2}Sample preparation}

\begin{figure*}[ht]
 \centering
 \includegraphics[width=14cm]{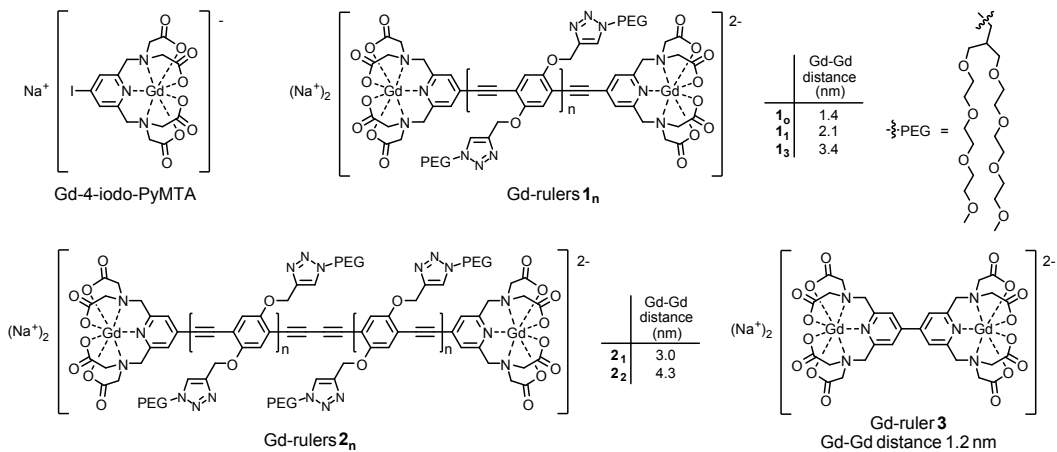}
 \caption{Chemical structures of Gd-4-iodo-PyMTA and of Gd-rulers $\mathbf{1_{n}}$, $\mathbf{2_{n}}$, and $\mathbf{3}$ used in this study. Gd-4-iodo-PyMTA served as a reference for the intrinsic CW EPR lineshape of the spin label in the absence of dipolar broadening. The listed Gd-Gd distances are the calculated most probable distances at 173 K, i.e. the glass transition of a 60:40 (v:v) mixture of \ce{D2O} and \ce{glycerol-d8} used as the matrix for the EPR experiments at 30 K.}
 \label{fgr:Fig1}
\end{figure*}

The syntheses of the Gd-rulers $\mathbf{1_{1}}$ and $\mathbf{1_{3}}$\cite{Qi} and of Gd-NO3pic\cite{Gateau2003} have been published. The syntheses of the Gd-rulers $\mathbf{2_{1}}$ and $\mathbf{2_{2}}$ and of Gd-4-iodo-PyMTA are reported in the supplementary information. The syntheses of Gd-rulers $\mathbf{1_{0}}$ and $\mathbf{3}$ will be reported elsewhere. Gd-DOTAM was purchased from Macrocyclics.

For sample preparation, stock solutions of Gd-rulers and Gd-4-iodo-PyMTA in \ce{D2O} were used. These solutions contained additional compounds remaining as a result of the synthesis, as detailed in Table S1. For measurements at cryogenic temperatures, stock solutions of Gd-rulers and Gd-4-iodo-PyMTA were diluted with a 60:40 (v:v) mixture of \ce{D2O} and \ce{glycerol-d8} (Cambridge Isotopes Laboratories, Inc.) to a concentration of 300 $\mu$M. Stock solutions of Gd-NO3Pic and Gd-DOTAM in \ce{D2O} were also diluted with a 60:40 (v:v) mixture of \ce{D2O} and \ce{glycerol-d8} to a concentration of 300 $\mu$M. For measurements near room temperature, the Gd-rulers were immobilized in dehydrated amorphous trehalose.\cite{Meyer2015} For this purpose, stock solutions of Gd-rulers and Gd-4-iodo-PyMTA were diluted with a 0.2 M solution of trehalose dihydrate (Sigma Aldrich) in \ce{D2O} (Cambridge Isotopes Laboratories, Inc.). The resulting samples had a mole ratio of 40:1 trehalose:Gd-ruler and 40:1 trehalose:Gd-4-iodo-PyMTA. These mixtures were then deposited onto a glass slide, allowed to dry for several days under a flow of dry nitrogen at room temperature, and finally placed under vacuum for at least 24 hours before measurement. The fragile solid was removed from the glass slide, crumbled into a powder, and transferred to a Teflon sample cup for measurements.

\subsection{\label{sec:level2}240 GHz CW EPR measurements}

CW EPR measurements were carried out on a home-built pulsed and CW EPR spectrometer operating at 240 GHz and equipped with a low power solid state source, quasi-optical bridge, and induction-mode superheterodyne detection, as has been described in detail elsewhere. \cite{Takahashi2012,Edwards2013,Edwards2013a} A solid-state source, which multiplies a 15 GHz synthesizer 16x to achieve 240 GHz, produces CW power of 50 mW (Virginia Diodes, Inc.). Samples of 8 - 10 $\mu$L volume were placed into a Teflon sample cup with a $\sim$ 3.5 mm inner diameter and $\sim$ 5 mm in height. The sample was backed by a mirror and mounted within a modulation coil at the end of an overmoded waveguide (Thomas Keating Ltd). This assembly was loaded into a continuous flow cryostat (Janis Research Company) mounted in the room-temperature bore of the magnet. Liquid samples were loaded into the $\sim$ 220 K precooled cryostat and then cooled to the desired measurement temperature at a rate of 3 - 4 K per minute. Incident microwave power was controlled by voltage-controlled attenuation of the source and a pair of wiregrid polarizers, and was on the order of $\mu$W at the sample. CW EPR measurements were carried out using field modulation at 20 kHz with 0.1 - 0.3 mT modulation amplitude to keep modulation at $<$ 1/5 of the linewidth of the central transition of \ce{Gd^{3+}} to avoid artificial line broadening.\cite{Eaton2010} No resonant cavity was used. The main coil of a sweepable 0 - 12.5 T superconducting magnet (Oxford Instruments) was used to carry out measurements at a sweep rate of 0.1 mT/second centered about 8.6 T magnetic field. Linearity of the magnetic field over the sweep range was verified with independent measurements using \ce{Mn^{2+}} in \ce{MgO} as a field standard (not shown).\cite{Krinichnyi1991} Superheterodyne detection was achieved using a Schottky subharmonic mixer (Virginia Diodes, Inc.) to mix the 240 GHz signal down to 10 GHz. A home-built intermediate-frequency stage then amplified and mixed this signal down to baseband. The resulting signal was measured in quadrature with a pair of lock-in amplifiers (Stanford Research Systems). 

The phase of the CW EPR spectra was set in post-processing. Each lock-in amplifier produces two components which are at angles $\phi$ and $\phi + 90^{\circ}$ with respect to a reference signal taken from the field modulation frequency. The angle $\phi$ for each lock-in amplifier was chosen such that the quadrature signal was minimized using a least-squares criterion, giving maximum signal-to-noise ratio in the in-phase signal.\cite{Auteri1988} The resulting in-phase signals after this phase nulling procedure were the real and imaginary components of the CW EPR spectra. These real and imaginary components were then rephased to determine the derivative CW EPR lineshape by equalizing the positive and negative peaks of the derivative lineshape.\cite{Edwards2013}

Peak-to-peak linewidths were determined by the following procedure, as necessitated by the broad and indistinct peaks observed with the shorter Gd-rulers (e.g. Fig. 2, D-F). First, the approximate locations of the positive and negative peaks in the CW EPR spectrum were found by looking for zero-crossings in the smoothed first derivative of the CW EPR spectra that exceed an input amplitude threshold chosen to exclude any zero-crossings that may be present in the baseline. The region around these approximate peak locations in the unsmoothed CW EPR spectra were then fit to a third-order polynomial and the extrema of the cubic fits taken as the location of the positive and negative peak positions in the experimental CW EPR spectrum. This procedure was repeated for several measurements of each compound to assess the degree of reproducibility of the measured lineshape and the error in peak to peak linewidths as determined by the above fitting procedure. The reported peak-to-peak linewidths of each compound were computed by taking a weighted average the peak-to-peak linewidths determined from many repeated EPR measurements of a particular compound. The weighting factor was taken to be the resolution with which the peak position could be determined, as given by the inverse of the standard deviation of the residual of the cubic fit to the positive and negative peak in each scan. The error in peak-to-peak broadening (included in Fig. 3) were computed by propagating the errors associated with the weighted average of the peak-to-peak linewidths of each Gd-ruler and Gd-4-iodo-PyTMA.\cite{Taylor1997}

\section{\label{sec:level1}Results and analysis}

A series of six water-soluble Gd-rulers $\mathbf{1_{n}}$, $\mathbf{2_{n}}$, and $\mathbf{3}$ (Fig. 1) spanning a Gd-Gd distance range of 1.2 nm to 4.3 nm were used. The Gd-rulers are based on oligo(\textit{para}-phenyleneethynylene)s (oligoPPEs) as the spacers and the Gd-PyMTA complex as the spin label that is connected to the oligoPPEs via its pyridine ring. The PEG side chains confer water solubility to the Gd-rulers.\cite{Qi} The length and flexibility of the oligoPPE spacers are well known from PDS of nitroxide rulers at X-band,\cite{Godt2006,Jeschke2010} and Gd-rulers of the type used in this study have previously been examined by PDS at Q-band\cite{Doll2015a} and W-band.\cite{Razzaghi2014,Dalaloyan2015} Based on experimental data\cite{Godt2006,Jeschke2010} and applying the worm-like chain model as reported earlier,\cite{Dalaloyan2015} the most probable Gd-Gd distances were calculated for the temperatures 173 K, 215 K, and 288 K. The resuls are summarized in Table S2. In Figure 1, the most probable Gd-Gd distances calculated for a temperature of 173 K are given. These distances are also attached as an easy identifier to the compound numbers throughout the text (e.g. Gd-ruler $\mathbf{1_{0}}$ (1.4 nm)). The temperature of 173 K corresponds to the glass transition of a 60:40 (v:v) mixture of \ce{D2O} and \ce{glycerol-d8} which was used as the matrix for the experiments at 30 K and therefore to the temperature at which the shape of the Gd-rulers becomes frozen upon cooling to 30 K.  

Samples for measurements at 30 K and 215 K were prepared in deuterated solvents to minimize the broadening from hyperfine interactions with neighboring water protons. A high fraction of glycerol was used to ensure good glass formation upon freezing. A sample concentration of 300 $\mu$M was chosen to avoid contributions to the lineshape from refractive broadening\cite{Edwards2013} and to ensure that the average intermolecular separation was sufficiently large so that only intramolecular dipolar interactions are observed in the lineshape. Because the magnitude of the dipolar coupling is here determined by a broadening of the CW EPR lineshape, it is necessary to have a measure of the intrinsic lineshape of the spin label in the absence of dipolar broadening.\cite{Rabenstein1995,Steinhoff1997} To determine this intrinsic lineshape, Gd-4-iodo-PyMTA, a \ce{Gd^{3+}} complex closely resembling the spin label of the Gd-rulers, (Fig. 1) was included in the study. We assume a negligible effect of the type of substituent at the pyridine ring on the lineshape.

\begin{figure*}
 \centering
 \includegraphics[width=17.1cm]{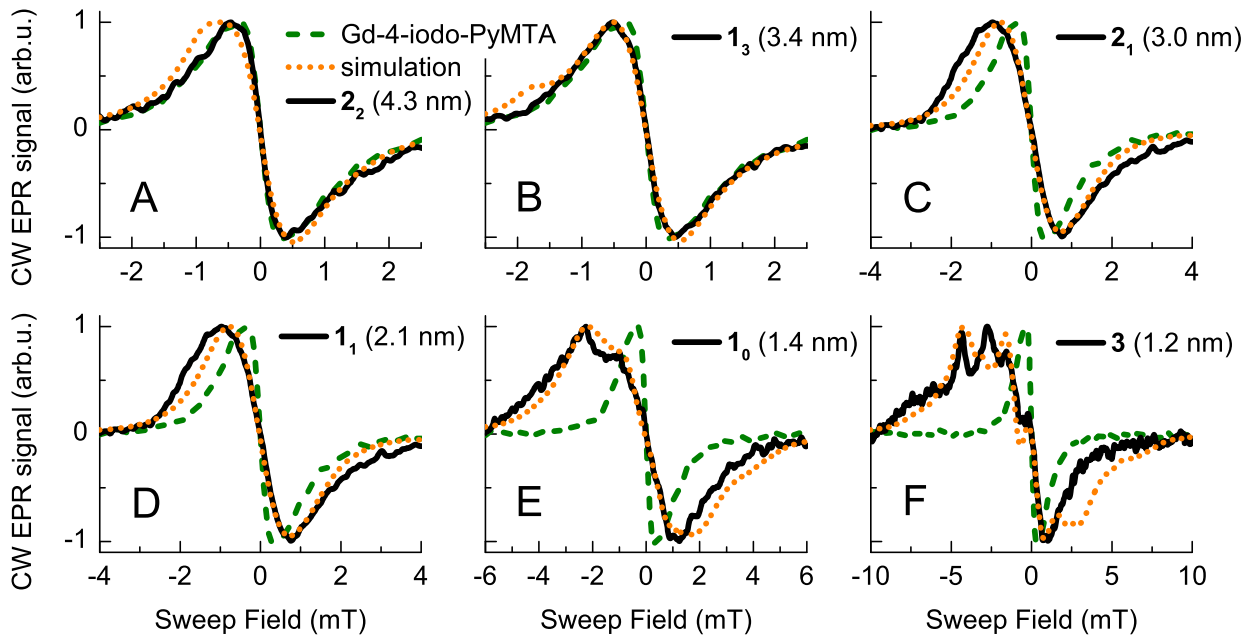}
 \caption{Lineshapes of the central transition of the Gd-rulers 1n, 2n, and 3 (A-F, solid black curves) in \ce{D2O}/\ce{glycerol-d8} measured by CW EPR at 240 GHz and 30 K. In (A-F), the measured line of Gd-4-iodo-PyMTA is overlaid in dashed green. In dotted orange, the simulated line resulting from simulations with a simple model consisting of an $S = 1/2$ spin whose CW EPR line is broadened through electron-electron dipole interactions with a neighboring $S = 7/2$ spin is plotted. In these simulations, the magnitude of the dipolar interaction was taken to be that corresponding to the calculated most probable distance for each Gd-ruler (Table S2).}
 \label{fgr:Fig2}
\end{figure*}

We begin by presenting the CW EPR lineshape of the central $|-1/2\rangle\rightarrow|1/2\rangle$ transition of the Gd-rulers and Gd-4-iodo-PyMTA measured at 240 GHz and a temperature of 30 K. For Gd-4-iodo-PyMTA, the central transition is a single peak. The peak-to-peak linewidth, taken as the separation of the positive and negative peaks in the derivative lineshape, was measured to be $\sim$ 0.77 mT for Gd-4-iodo-PyMTA (Fig. S1). This is broader than the $\sim$ 0.55 mT intrinsic linewidth at 240 GHz and 10 K of \ce{GdCl3} in \ce{D2O}/\ce{glycerol-d8},\cite{Edwards2013} but is still sufficiently narrow to detect broadening of the CW EPR lineshape by dipolar coupling with neighboring \ce{Gd^{3+}} ions. The central transition of Gd-4-iodo-PyMTA and the Gd-rulers was additionally measured by echo-detected (ED) EPR at W-band (95 GHz) and 10 K (see Fig. S3A). At this reduced field, the peak-to-peak linewidth of Gd-4-iodo-PyMTA is significantly increased to $\sim$ 1.6 mT. This increase in peak-to-peak linewidth from 240 GHz (8.6 T) to 95 GHz (3.4 T) is consistent with the expectation that the linewidth of the central transition scales as $D^{2}/g\mu_{B}B_{0}$.

The 240 GHz CW EPR spectra of the Gd-rulers at 30 K are shown overlaid with the spectrum of Gd-4-iodo-PyMTA in Fig. 2A-F. They reveal that the lineshape of the Gd-ruler $\mathbf{2_{2}}$ (4.3 nm) is nearly indistinguishable from that of Gd-4-iodo-PyMTA (Fig. 2A). Obviously, at a distance of 4.3 nm, broadening from the intramolecular dipolar coupling is smaller than the intrinsic linewidth of Gd-4-iodo-PyMTA. For the Gd-ruler $\mathbf{1_{3}}$ (3.4 nm), a slight broadening of the peak-to-peak linewidth to $\sim$ 0.97 mT is observed (Fig. 2B). This is consistent with the estimation of Edwards, et al. of a Gd-Gd distance sensitivity up to $\sim$ 3.8 nm.\cite{Edwards2013} Given that the intrinsic linewidth of Gd-4-iodo-PyMTA is $\sim$ 40\% larger than that of \ce{GdCl3} in \ce{D2O}/\ce{glycerol-d8}, the ability to resolve dipolar broadening of the CW EPR lineshape at extended distances using Gd-PyMTA as the spin label is expected to be somewhat reduced. In the case of Gd-ruler $\mathbf{2_{1}}$ (3.0 nm), the peak-to-peak linewidth is 1.15 mT - an increase of nearly 50\% with respect to the intrinsic linewidth given by Gd-4-iodo-PyMTA (Fig. 2C). For Gd-rulers $\mathbf{1_{1}}$ (2.1 nm) and $\mathbf{1_{0}}$ (1.4 nm) the dipolar broadening is even more dramatic with a peak-to-peak linewidth of 1.75 mT (Fig. 2D) and 3.47 mT (Fig. 2E), respectively. For the shortest Gd-ruler $\mathbf{3}$ (1.2 nm) we measure a further increased peak-to-peak linewidth of 5.22 mT (Fig. 2F), where the positive peak position was determined by fitting a cubic polynomial to the full low-field half of the spectrum shown in Fig. 2F. Additionally, it may be seen that for Gd-ruler $\mathbf{1_{1}}$ (2.1 nm) the lineshape has become distinctly asymmetric, with the low-field side becoming wider than the high-field side. As the interspin distance is decreased, this asymmetry in the lineshape becomes even more pronounced. For Gd-rulers $\mathbf{1_{0}}$ (1.4 nm) and $\mathbf{3}$ (1.2 nm) a number of additional peaks are seen to appear in the low-field side of the lineshape, with the high-field side remaining virtually unchanged between these distances.

To investigate the correlation between CW EPR line broadening and Gd-Gd distance, the change in linewidth resulting from intramolecular dipolar coupling of the two \ce{Gd^{3+}} ions of the Gd-rulers was expressed as the peak-to-peak broadening, taken to be the difference between the peak-to-peak linewidth of a Gd-ruler and the intrinsic linewidth given by Gd-4-iodo-PyMTA. The plot of the peak-to-peak broadening versus the calculated most probable Gd-Gd distances (Fig. 3) shows that the decrease of the peak-to-peak broadening with increasing distance $r$ scales as $1/r^{3}$. Such a correlation is expected from the dipolar coupling term in the effective spin Hamiltonian, as outlined in the next section.

A simple model to describe the two-spin dipolar coupled Gd-Gd system was developed to probe the dominant contributions to the lineshapes observed in the CW EPR spectra of the Gd-rulers. We take as a starting point the general effective spin Hamiltonian for a pair of two interacting \ce{Gd^{3+}} ions A and B which is given by\cite{Schweiger2001,Abragam1970}

\begin{equation}
     {\cal H} = \sum_{i=A,B}[g_{i}\mu_{B}\overrightarrow{B}_{0}\hat{S}_{Zi} + \hbar\cdot \mathbf{A}_{i}\cdot \hat{I}_{i} + \hbar\hat{S}_{A}\cdot \mathbf{D}\cdot\hat{S}_{B}] + \hbar\hat{S}_{A}\cdot\mathbf{T}\cdot\hat{S}_{B}
\end{equation}

The first term gives the contribution from the isotropic electron Zeeman interaction, where $\mu_{B}$ is the Bohr magneton, $B_{0}$ is the external applied magnetic field, and $g\sim$1.992. The second term is the contribution from hyperfine coupling, which arises from the \ce{^{155}Gd} and \ce{^{157}Gd} isotopes of gadolinium with a 30\% combined natural abundance.  However, the resulting hyperfine coupling is small - on the order of 16 MHz. For Gd-PyMTA as well as for most Gd complexes it is unresolved, contributing only a small amount to the intrinsic linewidth.\cite{Borel2006} The third term is the zero-field splitting (ZFS) interaction, whose magnitude is small compared to the electron Zeeman interaction and can be treated by perturbation theory. To first order in perturbation theory, the linewidth of the central $|-1/2\rangle\rightarrow|1/2\rangle$ transition of \ce{Gd^{3+}} is not affected by the ZFS interaction; the remaining higher-order transitions scale linearly with the second-order axial ZFS parameter $D$. To second order, the linewidth of the $|-1/2\rangle\rightarrow|1/2\rangle$ transition scales with $D^{2}/g\mu_{B}B_{0}$. This results in the characteristic high-field \ce{Gd^{3+}} EPR spectrum consisting of an intense narrow central peak arising from the $|-1/2\rangle\rightarrow|1/2\rangle$ transition and a broad featureless background due to all other transitions which are smeared out by the large distribution of ZFS parameters. The final term in Eqn. (1) is the electron spin dipole-dipole coupling, where $\mathbf{T}$ is a tensor describing the total interaction between two electron spins. The dominant contribution to this term comes from the secular part of the dipolar interaction, which in the point dipole approximation is given by

\begin{equation}
     {\cal H}^{secular}_{dd} = \omega_{dd}^{0}S_{Z}^{A}S_{Z}^{B}(3cos^{2}\theta - 1)
\end{equation}

where the magnitude of the dipolar coupling is given by

\begin{equation}
     \omega_{dd}^{0} = \frac{\mu_{0}}{4\pi}\frac{g_{A}g_{B}\mu_{B}^{2}}{\hbar}\frac{1}{\overrightarrow{r}_{AB}^{3}}
\end{equation}

$\theta$ is the angle between the interspin vector $r_{AB}$, $g_{A}$ and $g_{B}$ are the $g$ factors of the two spins, and $\mu_{B}$ is the Bohr magneton. For short interspin distances and small ZFS the pseudo-secular part of the dipolar interaction

\begin{equation}
     {\cal H}^{pseudosecular}_{dd} = -\frac{\omega_{dd}^{0}}{4}(S_{+}^{A}S_{-}^{B} + S_{-}^{A}S_{+}^{B})(3cos^{2}\theta - 1)
\end{equation}
will also contribute significantly to the measured central transition.

Direct simulation of the full effective spin Hamiltonian to extract the interspin separation $r_{AB}$ from the CW EPR lineshape would be computationally intensive, and furthermore would require an accurate \textit{a priori} knowledge of the magnitude and distribution of ZFS parameters for Gd-PyMTA. Therefore, we make the following simplifying assumptions in our simulation of the Gd-rulers. First, we note that the observed CW EPR lineshape of the central transition at 240 GHz is dominated by the $|-1/2\rangle\rightarrow|1/2\rangle$ transition, the other transitions being sufficiently smeared out by the broad distribution of ZFS parameters in a glassy sample that they are not explicitly resolved in this measurement. Therefore, we approximate the Gd-rulers as an $S=1/2$ spin whose CW EPR line is broadened through dipolar interactions with a neighboring $S=7/2$ spin. Next we consider that the ZFS interaction affects the central transition only to second-order, scaling as $\sim D^{2}/g\mu_{B}B_{0}$. At 240 GHz (8.6 T), the ZFS is much smaller than the static magnetic field ($D \sim$ 1150 MHz ($\sim$ 41 mT) for Gd-PyMTA),\cite{Dalaloyan2015} with the result that its primary contribution to the lineshape of the central transition of \ce{Gd^{3+}} is captured by determining the intrinsic lineshape of Gd-PyMTA. Hence, we introduce a further simplifying assumption, namely that the ZFS interaction can be accounted for with an artificial broadening imposed on the line of the $S=1/2$ spin. 

Based on the above mentioned simplifications to the effective spin Hamiltonian, the simulations were carried out as follows in Matlab (Mathworks 2014a) using the relevant functions from the EasySpin package\cite{Stoll2006} (version 5.0.16) with exact diagonalization. The simulated spin system consisted of an $S=1/2$ spin and an $S=7/2$ spin. A temperature of 30 K was chosen in the simulations to match the experimental temperature and to accurately account for the Boltzmann distribution of the $S=7/2$ spin populations. The lineshape of Gd-4-iodo-PyMTA was reproduced by introducing an artificial broadening in the simulations as a strain on the isotropic \textit{g}-value of the $S=1/2$ spin. This \textit{g}-strain was taken to be a Lorentzian distribution of \textit{g}-values centered at \textit{g} = 1.992 with a FWHM of 0.00028, chosen such that the resulting simulated lineshape matches as closely as possible the measured CW EPR lineshape and 0.77 mT linewidth of the central peak of Gd-4-iodo-PyMTA  (Fig. S1). The isotropic \textit{g}-value of the $S=7/2$ spin was taken to be also \textit{g} = 1.992. A small Voigtian convolutional line broadening (0.2 mT Gaussian + 0.5 mT Lorentzian) was included so that the simulated derivative produced a smooth line. The \textit{g}-strain on the $S=1/2$ spin and the convolutional line broadening used to reproduce the lineshape of Gd-4-iodo-PyMTA represent the only free parameters in these simulations. The broadening of the line of the Gd-rulers was achieved in simulation by introducing an isotropic electron-electron dipolar coupling in EasySpin, which uses the full interaction tensor including contributions from the secular (Eqn. 2), and pseudo-secular (Eqn. 4) terms of the dipolar coupling. The magnitude of the electron-electron dipolar coupling $\omega_{dd}^{0}$ (Eqn. 3) used in these simulations correspond to that at the calculated most probable Gd-Gd distance for a Gd-ruler at 173 K (Table S2).

The results of these simulations are overlaid on the spectra of the Gd-rulers in Fig. 2. Remarkably, given the many simplifying assumptions these simulations are based on, the measured lineshape of the Gd-rulers are well described by the simulations. Particularly in the case of the short Gd-rulers $\mathbf{1_{0}}$ (1.4 nm) and $\mathbf{3}$ (1.2 nm), many of the details of the measured lineshape - including approximate positions and amplitudes of the splittings resulting from interaction with an $S = 7/2$ spin - are reproduced. Furthermore, the dramatic broadening and complex lineshape observed for Gd-Gd distances of 1.2 nm - 1.4 nm (Figs. 2 E-F) suggest that sub-\AA ngstrom Gd-Gd distance discrimination within this range is possible with Gd-based spin labels. While not quantitative at this level, these simulations do validate the assumption that the observed broadening of the CW EPR line of the Gd-rulers is arising primarily from the electron-electron dipolar interaction between two Gd complexes and allow for the determination of the Gd-Gd distance from the magnitude of the dipolar broadening.

\begin{figure}
\centering
  \includegraphics[width=8.3cm]{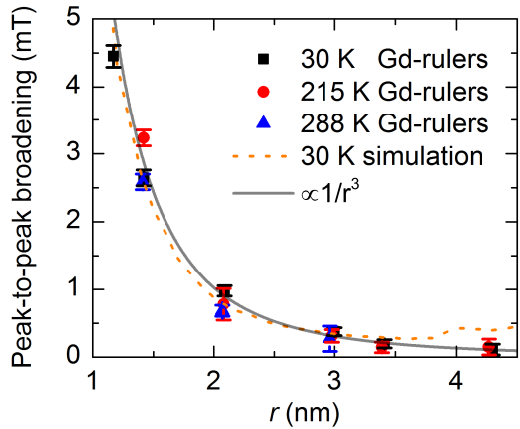}
  \caption{Dipolar broadening in the spectra of the Gd-rulers at 30 K, 215 K, and 288 K plotted as a function of the calculated most probable Gd-Gd distances at 173 K (glass transition temperature of the matrix), 215 K, and 288 K, respectively. As a measure for the dipolar broadening, the peak-to-peak broadening, which is the peak-to-peak linewidths of the Gd-rulers less the intrinsic linewidth of Gd-4-iodo-PyMTA at the same temperature is taken. For all temperatures, the dipolar broadening falls off as $1/r^{3}$. Peak-to-peak broadening of the simulated line follows a similar trend up to $\sim$ 3.4 nm.}
  \label{fgr:Fig3}
\end{figure}

Next, we present CW EPR measurements performed on identically prepared samples at 215 K, i.e. at a temperature at or near protein dynamical transition,\cite{Doster2010} at which the sample is still sufficiently viscous that the dipolar coupling is not completely averaged out by molecular tumbling.\cite{Gonzalez2011} The intrinsic linewidth from Gd-4-iodo-PyMTA was seen to increase from 0.77 mT at 30 K to 1.0 mT at 215 K (Fig. S2A). This increasing linewidth with increasing temperature was unexpected based on our simulations and prior work with \ce{GdCl3}\cite{Edwards2013} and will require further investigation. Nevertheless, the lineshape of the Gd-rulers $\mathbf{1_{n}}$ and $\mathbf{2_{n}}$ at 215 K show a comparable correlation of broadening with the most probable Gd-Gd distance as seen at 30 K (Fig. 3, Fig. S2). At this temperature, it is possible to resolve a change in linewidth of the Gd-ruler $\mathbf{2_{1}}$ (3.0 nm) with respect to the Gd-4-iodo-PyMTA lineshape, whereas for Gd-rulers $\mathbf{1_{3}}$ (3.4 nm) and $\mathbf{2_{2}}$ (4.3 nm), dipolar coupling effects fall within the noise of the measurement at 215 K. This reduction in maximum resolvable distance at 215 K to approximately 3.0 nm from 3.4 nm at 30 K is not surprising, given the increase in the intrinsic linewidth and reduction in signal to noise ratio with elevated temperature, as well as partially decreased dipolar broadening for elevated temperatures as a result of changing spin populations with temperature.

To test the viability of lineshape analysis for distance determination at room temperature, CW EPR spectra were also recorded at 288 K of Gd-4-iodo-PyMTA and the Gd-rulers immobilized in glassy trehalose (Fig. S2), inspired by the PDS studies performed by Eaton and coworkers.\cite{Meyer2015} The trehalose forms an amorphous matrix which is solid at 288 K and thus inhibits averaging out of the dipolar interaction by molecular tumbling.\cite{Meyer2015,Palazzo2002,Simperler2007} The molar ratios of trehalose to the Gd-rulers were chosen such as to avoid line broadening due to intermolecular interactions if the Gd-rulers are homogeneously distributed in the matrix. For the Gd-rulers $\mathbf{1_{3}}$ (3.4 nm) and $\mathbf{2_{2}}$ (4.3 nm) it was not possible to record a CW EPR spectrum at 288 K with sufficient signal to noise ratio (SNR) to determine the peak-to-peak linewidth. This was attributed to difficulties in sample preparation and not to an intrinsic limitation of the experimental technique. The drying process was found to take a significantly longer time with the long Gd-rulers than with the short Gd-rulers, possibly caused by the increased content of the hydrophilc PEG side chains. If a partial demixing of the Gd-rulers from the matrix during the slow drying process of the aqueous mixtures of trehalose and the Gd-rulers occured, an increase in local concentration of Gd-rulers cannot be ruled out. Gd-ruler $\mathbf{3}$ (1.2 nm) was not available at the time of these experiments. Measurements with the other Gd-rulers and Gd-4-iodo-PyMTA revealed again an increased linewidth with increased temperature, with the linewidth of Gd-4-iodo-PyMTA measured to be 1.39 mT at 288 K (Fig. S2A). Nevertheless, even at this elevated temperature and despite significant changes of the sample environment, gratifyingly the $1/r^{3}$ correlation of peak-to-peak line broadening with the calculated most probable Gd-Gd distance of the Gd-rulers at 288 K is observed (Fig. 3, Fig. S2). At 288 K, peak-to-peak broadening of the CW EPR lineshape is clearly resolved for Gd-rulers $\mathbf{1_{0}}$ (1.4 nm) and $\mathbf{1_{1}}$ (2.1 nm), but cannot be unambiguously determined for a distance for $\mathbf{2_{1}}$ (3.0 nm). This reduction of distance sensitivity in comparison to the studies at 215 K is at least partially a result of the increased intrinsic linewidth of Gd-4-iodo-PyMTA at this temperature and dramatically decreased SNR when compared to frozen samples. Additionally, the possible demixing of the Gd-rulers from the trehalose matrix as discussed above could also contribute to the observed increase in linewidth and corresponding decrease in distance sensitivity.

\section{\label{sec:level1}Discussion}

With Gd-PyMTA, a Gd-based spin label, spin distances from 1.2 nm up to between 3.4 and 4.3 nm can be determined by lineshape analysis of a CW EPR spectrum of the central $|-1/2\rangle\rightarrow|1/2\rangle$ transition recorded at 30 K and 240 GHz. In comparison, the upper limit when using nitroxide-based radicals is about 2 nm at X-band ($\sim$ 10 GHz) frequencies.\cite{Hubbell1998} The origin of this increased distance sensitivity when using Gd-based spin labels measured at high frequencies is twofold. First, the narrow central line of the \ce{Gd^{3+}} spectrum acts as a much more sensitive probe of dipolar broadening effects when compared to the rather broad nitroxide line. Second, dipolar interactions between two \ce{Gd^{3+}} ions cause greater line broadening for a given distance than would be seen for two nitroxides with $S=1/2$ because the $S=7/2$ of \ce{Gd^{3+}} allows for 8 possible spin states of the neighboring spin. These additional states contribute to the zero-order dipolar interaction felt by the other spin, causing a larger shift in the local dipolar field and therefore a larger broadening as compared to broadening caused by a proximal $S=1/2$ spin label.\cite{Abragam1970}

The results of the lineshape analysis of CW EPR spectra recorded at 240 GHz of Gd-4-iodo-PyMTA and the Gd-rulers $\mathbf{1_{n}}$, $\mathbf{2_{n}}$, and $\mathbf{3}$ place the upper distance limit for measuring significant dipolar broadening when employing Gd-PyMTA as spin labels between 3.4 nm and 4.3 nm at 30 K. However, this is not expected to represent an absolute upper limit for this technique, as the distance sensitivity of this measurement is highly dependent on the intrinsic linewidth of the chosen Gd-complex. For Gd-4-iodo-PyMTA, the peak-to-peak CW EPR linewidth was found to be $\sim$ 0.77 mT at 240 GHz and 30 K (Fig. S1), while complexes such as Gd-DOTAM and Gd-NO3Pic, which have very narrow EPR linewidths already at X- and W-band frequencies,\cite{Borel2006} have linewidths of 0.53 mT and 0.45 mT at 240 GHz, respectively (Fig. S4). This corresponds to even narrower EPR lines than observed with \ce{GdCl3} in \ce{D2O}/\ce{glycerol-d8}, allowing us to project that the distance sensitivity by CW EPR lineshape analysis can be further improved by an appropriate choice of spin label. If, conservatively, we assume that a 25\% increase of the intrinsic peak-to-peak linewidth can be clearly resolved as line broadening (as in Gd-ruler $\mathbf{1_{3}}$ (3.4 nm), Fig. 3), then assuming a $1/r^{3}$ dependence of the broadening on distance, a $\sim$ 4.0 nm distance should be determinable for a spin label with a 0.5 mT intrinsic linewidth. However, it remains to be seen, whether the narrow linewidth is retained when these Gd complexes are used as spin labels. To employ the Gd complexes as spin labels, they need to be chemically altered which may alter their intrinsic linewidth. Furthermore, non-covalent interaction with the compound to which the spin label is bound, such as proteins or lipid systems, will probably impact the linewidth. 

CW EPR is not fundamentally limited to 240 GHz frequencies. Microwave sources and instrumentation suitable for EPR at higher frequencies are becoming increasingly available, with several CW EPR studies demonstrated at 250 GHz frequencies and above.\cite{Lynch1988,Barra1990,Earle1996,Krzystek1999,Hassan2000} Because the intrinsic linewidth of the central $|-1/2\rangle\rightarrow|1/2\rangle$ transition of \ce{Gd^{3+}} scales with $D^{2}/g\mu_{B}B_{0}$, CW EPR lineshape analysis will directly benefit from studies at increasingly higher magnetic fields. This would be particularly advantageous for leveraging existing commercially available Gd-based spin labels, such as Gd-4MMDPA, whose intrinsic linewidth of $\sim$ 1.3 mT at 240 GHz is too broad for sensitive CW EPR distance measurements at 240 GHz.\cite{Edwards2013} The scaling of the linewidth of the central transition with magnetic field is demonstrated here with echo-detected EPR spectra recorded at W-band (95 GHz) and 10 K (Fig. S3A). Gratifyingly, the full-width at half maximum of the absorption lineshape remains approximately linear with $1/r^{3}$ for Gd-rulers with Gd-Gd distances $r$ ranging from 2.1 - 3.4 nm (Fig. S3B), but the magnitude of the broadening for a given distance is smaller than at 240 GHz, as expected.

The analysis presented here, wherein the peak-to-peak broadening of the CW EPR lineshape of the central transition of \ce{Gd^{3+}} is seen to follow a $1/r^{3}$ dependence, allows only for determination of the mean interspin distance. However, distance distributions often contain equally valuable information, e.g. on the relative frequency of conformers of a biomolecule or on aggregation resulting in multiple label-to-label distances, making it an important analysis target. CW EPR lineshape analysis should be sensitive to the interspin distance distribution, but extraction of this information will require a better understanding of the various factors contributing to the CW EPR lineshape. The simulations discussed above provide a first step in this direction to capture the primary contributions to the lineshape, but a more detailed analysis requires simulation of the full effective spin Hamiltonian of two interacting S = 7/2 spins, including contributions from the zero-field splitting term. Dalaloyan et al. have shown that below 3.4 nm and with small $D$ values - conditions under which the high-field CW EPR technique is most sensitive - careful inclusion of the effects of the zero-field splitting parameter $D$ on the pseudo-secular part of the dipolar interaction is crucial for extracting accurate distances and distance distributions from DEER of Gd-Gd systems.\cite{Dalaloyan2015,Cohen2016} If the ZFS parameters were independently determined for a particular \ce{Gd^{3+}} complex, simulations could be fit to the full CW EPR lineshape with the distance distribution as the only free parameter, rather than just extracting the peak-to-peak linewidth.

\section{\label{sec:level1}Conclusions}

Lineshape analysis of CW EPR spectra of rigid Gd-rulers with Gd-Gd distances ranging from 1.2 nm to 4.3 nm (Fig. 1) recorded at 240 GHz and 30 K demonstrates scaling of dipolar line broadening with $1/r^{3}$ and distance sensitivity from 1.2 nm up to $\sim$ 3.4 nm, when Gd-PyMTA is used as the spin label. The same $1/r^{3}$ dependence is observed at biologically relevant temperatures of 215 K and 288 K, with the upper distance limit reduced to $\sim$ 3.2 nm and $\sim$ 2.9 nm, respectively (Fig. 3). The origin of the reduction in the upper limit with increasing temperature is not yet understood. On the other hand, we have no indication that these limits are spectroscopically intrinsically defined. These results project that distance determination by lineshape analysis of CW EPR spectra recorded at very high frequencies using Gd complexes as spin labels is a highly useful technique for structure studies of complex biological systems where the application of PDS is challenging, or when measurements above the solvent glass transition temperature is desirable or necessary. Practical applications will benefit from employing \ce{Gd^{3+}} complexes with very narrow central EPR lines, independent of the local environment. The latter calls for \ce{Gd^{3+}} complexes with ligands filling all coordination sites of the \ce{Gd^{3+}} ion, being resistant to substitution of any coordinating functional group by moieties of the biomolecule, and being conformationally fixed as to keep the geometry independent of the environment.

\section{\label{sec:level1}Acknowledgments}
We thank C.B. Wilson for countless helpful discussions on this work, A. Dalaloyan for performing the W-band measurements, H.P. Dette and T. Koop for determination of the glass transition temperature of the mixture of \ce{D2O} and \ce{glycerol-d8}, A. Feintuch for helpful discussions on the simulations, and J. Wegner for preparing Gd-NO3Pic. This work was supported by funding from the NSF (Molecular and Cellular Biology grants MCB-1244651 and MCB-1617025), the NIH GM under Award Number 1R01GM116128-01, the Israel-USA BSF science foundation under grant number 2010130, and the DFG (SPP1601; GO 555/6-2).


\input{Gd_ruler_paper_for_ArXiV.bbl}

\end{document}

%% file: Gd_ruler_paper_for_ArXiV.bbl
\providecommand*{\mcitethebibliography}{\thebibliography}
\csname @ifundefined\endcsname{endmcitethebibliography}
{\let\endmcitethebibliography\endthebibliography}{}